\documentclass[10pt,twocolumn,letterpaper]{article}

\usepackage[final]{cvpr}   

\usepackage{graphicx}
\usepackage{amsmath}
\usepackage{amssymb}
\usepackage{booktabs}
\usepackage{array}
\usepackage{siunitx}
\usepackage{boldline}

\usepackage[pagebackref,breaklinks,colorlinks]{hyperref}

\usepackage[capitalize]{cleveref}
\crefname{section}{Sec.}{Secs.}
\Crefname{section}{Section}{Sections}
\Crefname{table}{Table}{Tables}
\crefname{table}{Tab.}{Tabs.}

\def\confName{CVPR}
\def\confYear{2022}

\begin{document}

\title{DeepSupp: Attention-Driven Correlation Pattern Analysis for Dynamic Time Series Support and Resistance Levels Identification}

\author{
  Boris Kriuk, Logic Ng, Zarif Al Hossain\\
  Hong Kong University of Science and Technology\\
  Clear Water Bay, Hong Kong\\
  \{bkriuk, lcngab, zahossain\}@connect.ust.hk
}

\maketitle

\begin{abstract}
  Support and resistance (SR) levels are central to technical analysis, guiding traders in entry, exit, and risk management. Despite widespread use, traditional SR identification methods often fail to adapt to the complexities of modern, volatile markets. Recent research has introduced machine learning techniques to address the following challenges, yet most focus on price prediction rather than structural level identification. This paper presents DeepSupp, a new deep learning approach for detecting financial support levels using multi-head attention mechanisms to analyze spatial correlations and market microstructure relationships. DeepSupp integrates advanced feature engineering, constructing dynamic correlation matrices that capture evolving market relationships, and employs an attention-based autoencoder for robust representation learning. The final support levels are extracted through unsupervised clustering, leveraging DBSCAN to identify significant price thresholds. Comprehensive evaluations on S\&P 500 tickers demonstrate that DeepSupp outperforms six baseline methods, achieving state-of-the-art performance across six financial metrics, including essential support accuracy and market regime sensitivity. With consistent results across diverse market conditions, DeepSupp addresses critical gaps in SR level detection, offering a scalable and reliable solution for modern financial analysis. Our approach highlights the potential of attention-based architectures to uncover nuanced market patterns and improve technical trading strategies.
\end{abstract}

\section{Introduction}

\begin{figure*}[t!] 
  \centering
  \includegraphics[width=\textwidth]{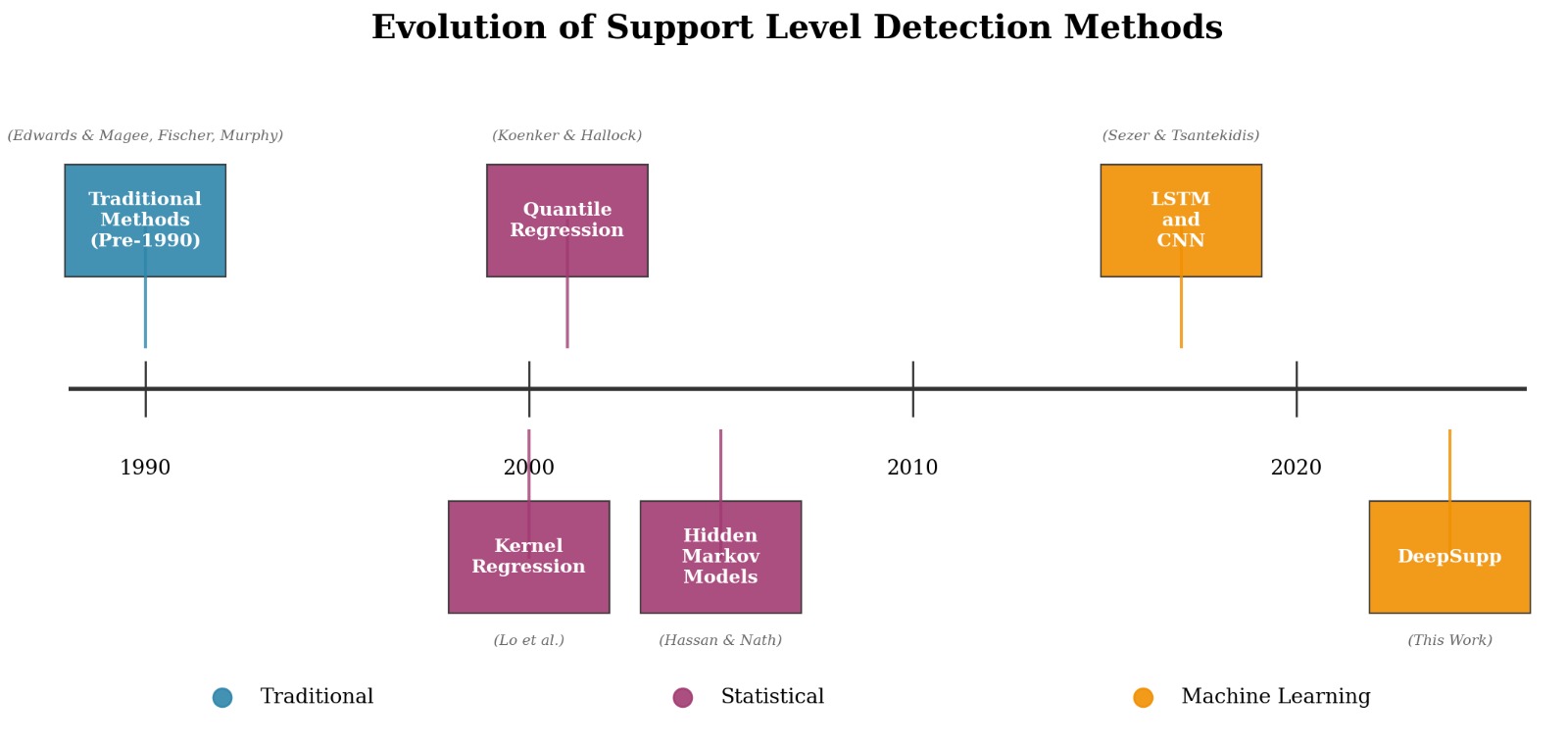} 
  \caption{Evolution of Support Level Detection Methods.}
  \label{fig:evolution_support_methods}
\end{figure*}

Support and resistance (SR) levels are foundational concepts in technical analysis, widely used by retail and institutional traders to guide entry, exit, and risk management decisions \cite{taylor1992use, zapranis2012identifying}. Such levels represent price thresholds where market participants expect a reversal or pause in price movement due to the aggregation of buy or sell pressure \cite{deangelis2016optimal}. When price approaches a support level, traders anticipate upward pressure from buyers; conversely, resistance levels are expected to trigger selling.

Empirical studies confirm the predictive relevance of SR levels \cite{osler2001currency, cadrin2014stock} found that SR levels published by financial firms had statistically significant effects on future price movements. Analysis of high-frequency tick data demonstrates that prices are more likely to bounce off than breach established SR levels \cite{garzarelli2014memory, yildirim2019evolutionary}. These findings align with the broader concept of SR levels acting as behavioral anchors in financial markets.

The practical utility of SR levels is further underscored by widespread adoption. Trading-range breakout strategies—based on SR levels—as among the most commonly used and effective technical trading rules \cite{brock1992simple}. Surveys show that technical analysis, which often centers around SR levels, is employed by 86\% of professional fund managers \cite{menkhoff2010use} and nearly a third of individual investors \cite{hoffmann2014technical}.

Traditional methodologies such as trend line analysis, moving averages, and Fibonacci retracements have long dominated practice, yet they lack the adaptability required in today's volatile markets. Recent advancements in computational methods have introduced more sophisticated approaches like quantile regression and hidden Markov models, although they still struggle with the non-stationary and multi-dimensional nature of financial time series.

This research makes three key contributions:

\begin{enumerate}
  \item We introduce DeepSupp, a new deep learning approach that utilizes multi-head attention mechanisms on spatial correlations to identify meaningful support levels.
  \item We provide a standardized comparison framework for evaluating support level detection methods using six fundamental financial metrics.
  \item We demonstrate that DeepSupp can capture subtle price-volume relationships that traditional methods miss, achieving state-of-the-art performance.
\end{enumerate}

\section{Related Works}

Early work in support level detection relied primarily on visual identification and simple moving averages. Fibonacci retracements gained popularity through the work of \cite{fischer1993fibonacci}, who connected these mathematical ratios to market psychology. \cite{williams1995trading, widner1998automated} introduced fractal theory to technical analysis, proposing that markets exhibit self-similar patterns across different timeframes.

Statistical approaches emerged with the work of \cite{lo2000foundations}, who applied kernel regression to identify significant support and resistance levels. Quantile regression methods were later proposed by \cite{koenker2005quantile, osler2000support} as a way to model the lower bounds of price movements. More recently, hidden Markov models have been applied to financial time series by \cite{hassan2005stock, martz2014systematic} to capture regime-switching behavior in markets.

The application of machine learning to technical analysis has grown significantly in recent years. \cite{dingli2017financial} provided a comprehensive survey of deep learning models in financial time series forecasting, while \cite{tsantekidis2017using} demonstrated the power of convolutional neural networks in capturing price patterns.

Despite these advancements, the specific application of deep learning to support level detection remains underexplored \cite{tsinaslanidis2023testing}. Most existing research focuses on price prediction rather than identifying structural levels in the market. This research gap motivates our development of DeepSupp, which specifically targets support level identification through attention-based mechanisms.

\section{Methodology}

This section introduces DeepSupp, a new deep learning method for identifying financial support levels that combines multi-head attention mechanisms with correlation analysis and clustering. Unlike traditional technical analysis approaches that rely on simple price patterns or basic statistical measures, DeepSupp captures complex non-linear temporal dependencies and market microstructure relationships.

\subsection{Feature Engineering and Data Preprocessing}

The foundation of DeepSupp lies in feature engineering that incorporates volume-weighted price action and market microstructure indicators. We begin with standard price and volume data and construct more complex features that capture market dynamics more effectively than raw price data alone.

The Volume Weighted Average Price (VWAP) provides a benchmark that reflects the true average price weighted by trading activity:
\begin{equation}
  \text{VWAP}_t = \frac{\sum_{i=1}^{t} P_i \cdot V_i}{\sum_{i=1}^{t} V_i}
\end{equation}

Price change weighted by volume captures the intensity of price movements, providing deeper understanding of the conviction behind price moves:
\begin{equation}
  \text{PriceChangeVolume}_t = \frac{P_t - P_{t-1}}{P_{t-1}} \cdot V_t
\end{equation}

The volume ratio compares current volume to recent average volume, identifying periods of unusual pattern activity:
\begin{equation}
  \text{VolumeRatio}_t = \frac{V_t}{\frac{1}{20}\sum_{i=t-19}^{t} V_i}
\end{equation}

These features are combined into a feature vector $\mathbf{F}_t = [\text{Close}_t, \text{VWAP}_t, \text{Volume}_t, \text{PriceChangeVolume}_t, \text{VolumeRatio}_t]$ and normalized using MinMax scaling to ensure numerical stability during training.

\begin{figure*}[h!]
  \centering
  \includegraphics[width=0.8\textwidth]{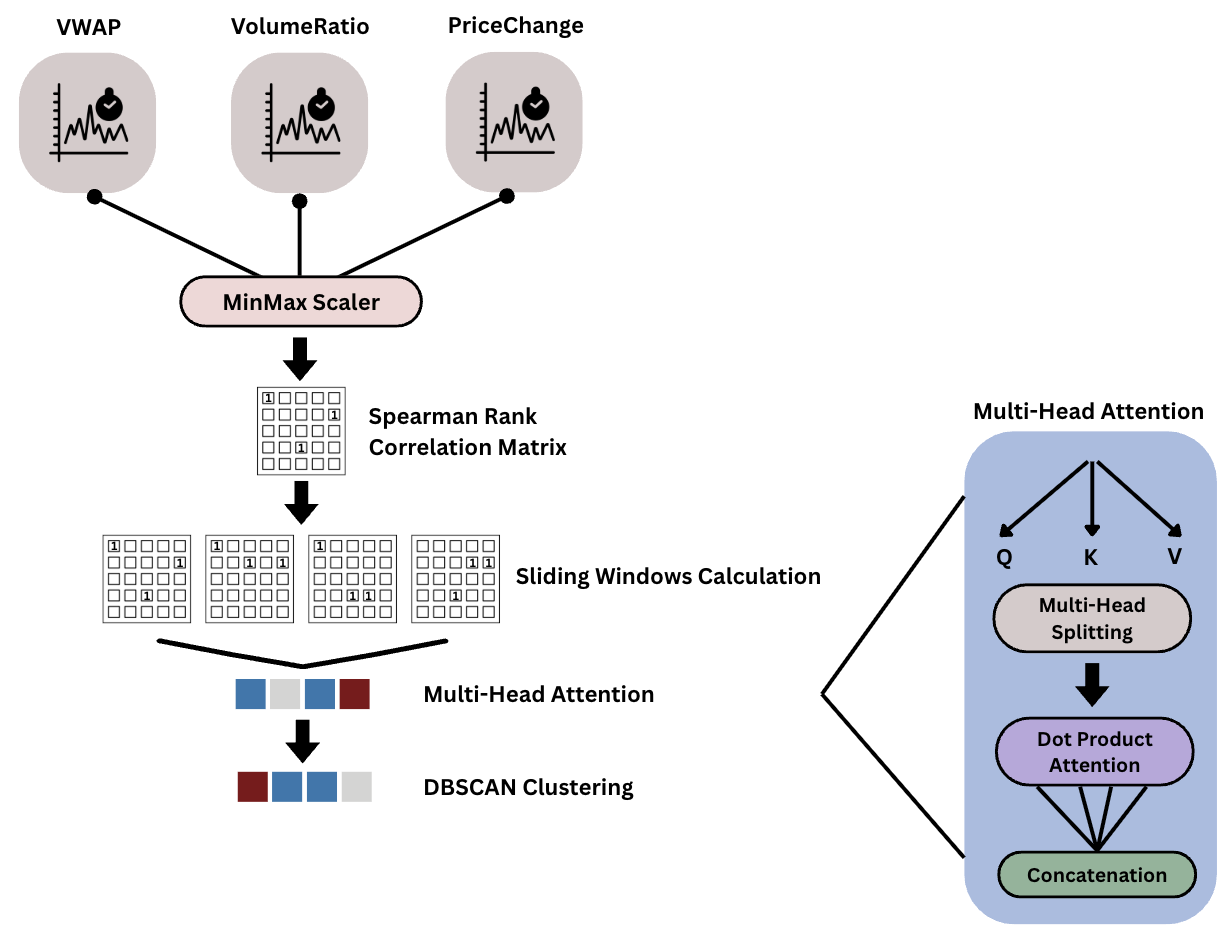}
  \caption{DeepSupp Architecture Overview: The methodology integrates feature engineering with volume-weighted indicators, dynamic correlation analysis using Spearman rank correlation, multi-head attention processing for parallel relationship analysis, autoencoder-based representation learning, and DBSCAN clustering for market regime identification.}
  \label{fig:deepsupp_architecture}
\end{figure*}

\subsection{Dynamic Correlation Analysis}

The correlation analysis module constructs dynamic correlation matrices that capture evolving relationships between market variables. Rather than assuming static relationships, our approach recognizes market correlations change over time, particularly during periods of market stress or rigid changes.

We employ Spearman rank correlation to capture non-linear relationships using sliding windows of length $n=32$:
\begin{equation}
  \rho_{ij}^{(t)} = 1 - \frac{6\sum_{k=1}^{n} d_{k}^2}{n(n^2-1)}
\end{equation}

where $d_k$ represents the difference between ranks of observations $i$ and $j$ at time $k$. The sliding window approach generates correlation matrices for overlapping time periods, creating a sequence of correlation snapshots that feed into the attention mechanism. Each correlation matrix is padded or trimmed to maintain consistent dimensions of $32 \times 32$ to ensure compatibility with the multi-head attention architecture. The correlation computation is performed across all feature pairs, resulting in symmetric matrices that capture both linear and non-linear dependencies between market variables.

\subsection{Multi-Head Attention Autoencoder}

The core idea relies on a specialized autoencoder that processes correlation matrices through multi-head attention mechanisms, as shown in Figure 2. A crucial design principle is leveraging the permutation invariance property of multi-head attention, which makes it naturally suited for processing correlation matrices. Since correlation matrices represent symmetric relationships between variables regardless of their ordering, the permutation invariance ensures that the attention mechanism focuses on the underlying market relationships rather than arbitrary feature sequences.

The architecture employs 4 attention heads with an embedding dimension of 32, ensuring the sequence length is divisible by the number of heads for optimal computational efficiency. This design allows each attention head to specialize in different types of market relationships while maintaining computational tractability.

The multi-head attention layer processes input correlation matrices $\mathbf{C}_t \in \mathbb{R}^{32 \times 32}$ through parallel attention heads. Each head learns different correlation patterns simultaneously, with query, key, and value transformations applied to capture diverse market relationship types. The permutation invariance property ensures that the attention weights remain consistent regardless of how the market variables are ordered in the correlation matrix, focusing purely on the strength and nature of the relationships. The attention mechanism computes scaled dot-product attention with residual connections and layer normalization to stabilize training.

\begin{table*}[t!]
  \centering
  \caption{Performance Comparison of Support Level Detection Methods}
  \label{tab:performance_results}
  \begin{tabular}{l@{\hspace{0.5cm}}c@{\hspace{0.5cm}}c@{\hspace{0.5cm}}c@{\hspace{0.5cm}}c@{\hspace{0.5cm}}c@{\hspace{0.5cm}}c@{\hspace{0.5cm}}c}
    \toprule
    \textbf{Method}     & \textbf{Overall}       & \textbf{Support}  & \textbf{Price}     & \textbf{Volume}       & \textbf{Market} & \textbf{Support}  & \textbf{Breakout} \\
                        & \textbf{Score}         & \textbf{Accuracy} & \textbf{Proximity} & \textbf{Confirmation} & \textbf{Regime} & \textbf{Duration} & \textbf{Recovery} \\
    \midrule
    \textbf{DeepSupp}   & \textbf{0.554 ± 0.039} & 0.483             & 0.759              & 0.349                 & \textbf{0.299}  & 0.846             & \textbf{0.800}    \\
    HMM                 & 0.550 ± 0.044          & 0.408             & \textbf{0.826}     & 0.348                 & 0.299           & \textbf{0.859}    & 0.800             \\
    Local Minima        & 0.507 ± 0.048          & \textbf{0.603}    & 0.362              & \textbf{0.351}        & 0.299           & 0.857             & 0.800             \\
    Fractal             & 0.478 ± 0.049          & 0.583             & 0.262              & 0.350                 & 0.299           & 0.831             & 0.800             \\
    Fibonacci           & 0.449 ± 0.044          & 0.570             & 0.137              & 0.349                 & 0.299           & 0.832             & 0.800             \\
    Moving Average      & 0.385 ± 0.081          & 0.311             & 0.168              & 0.349                 & 0.297           & 0.796             & 0.800             \\
    Quantile Regression & 0.336 ± 0.147          & 0.197             & 0.182              & 0.301                 & 0.297           & 0.744             & 0.684             \\
    \bottomrule
  \end{tabular}
\end{table*}

The encoder component consists of two linear layers with ReLU activations, compressing the attention-enhanced features from 32 dimensions to 16 dimensions. Such compression forces the model to learn most salient market characteristics by distilling the complex correlation patterns into essential market state representations. The decoder mirrors the encoder structure, reconstructing the original correlation matrices from the compressed embeddings using a symmetric architecture.

\subsection{Clustering-Based Support Level Extraction}

The final stage extracts support levels through unsupervised clustering of the learned embeddings. After training completion, embeddings are extracted from the encoder bottleneck layer, representing compressed market state information with dimensionality 16.

DBSCAN clustering is applied to the embedding space with parameters $\epsilon=0.1$ and minimum samples equal to 10\% of the dataset size. These parameters are chosen to identify dense regions in the embedding space while maintaining robustness to noise. The algorithm automatically determines the number of clusters based on the data density, avoiding the need for pre-specified cluster counts.

For each identified cluster $C_k$, the corresponding time indices are mapped back to the original price data. The median price level is computed for all time periods belonging to each cluster: $S_k = \text{median}\{P_t : t \in \text{TimeIndices}(C_k)\}$. The median is selected over the mean for robustness against price outliers and provides more stable support level estimates.

The final output consists of sorted support levels corresponding to different market regimes identified through the clustering process.

\section{Experiments}

We perform a comprehensive evaluation of DeepSupp against six baseline methods using S\&P 500 tickers with historical price and volume data over a 2 year time period. The evaluation employed six fundamental financial metrics as ground truth, providing a multi-dimensional assessment of support level detection performance. The baseline methods included Hidden Markov Models (HMM), Local Minima detection, Fractal analysis, Fibonacci retracement, Moving Average analysis, and Quantile Regression approaches.

Our evaluation framework incorporates six key performance indicators that capture different aspects of support level effectiveness. Support Accuracy measures how often prices bounce off predicted support levels with at least 1\% recovery. Price Proximity evaluates how closely support levels align with actual price percentiles from the 5th to 35th percentiles. Volume Confirmation assesses whether high volume accompanies support bounces, indicating institutional validation. Market Regime Sensitivity tests performance consistency across bull, bear, and sideways market conditions. Support Hold Duration measures how long support levels remain valid before breaking. False Breakout Rate evaluates recovery rate after brief support breaks within 3\% tolerance.

The overall score calculation employs a weighted combination of all six metrics to reflect their relative importance in practical trading applications. Support Accuracy receives the highest weight at 25\% due to its fundamental importance in bounce prediction, followed by Price Proximity and Volume Confirmation at 20\% each for their roles in price alignment and institutional validation. Market Regime Sensitivity and Support Hold Duration are weighted at 15\% each, reflecting their importance for strategy robustness and position management. False Breakout Rate receives the lowest weight at 5\%, as it represents a supplementary validation metric rather than a primary performance indicator.

DeepSupp achieved the highest overall score of 0.554 with the lowest variability at ±0.039, demonstrating superior and consistent performance across the weighted combination of all metrics evaluated on tickers. Our  method ranks first among all competing approaches, with HMM following closely at 0.550 ± 0.044 and Local Minima at 0.507 ± 0.048. The performance hierarchy continues with Fractal analysis at 0.478, Fibonacci retracement at 0.449, Moving Average at 0.385, and Quantile Regression at 0.336 with the highest variability of ±0.147.

The results reveal distinct category leaders across different performance dimensions. Local Minima detection achieved the highest support accuracy at 0.603, indicating superior performance in identifying immediate price bounce points. HMM demonstrated the best price proximity at 0.826, suggesting exceptional alignment with statistical price distributions. Local Minima also led in volume confirmation at 0.351, while HMM achieved the highest support duration stability at 0.859.

DeepSupp's competitive advantage lies in its balanced performance across multiple metrics combined with the lowest performance variability. While not achieving the top score in every individual category, DeepSupp demonstrates exceptional consistency with a standard deviation of only 0.039 compared to the significantly higher variability of traditional methods. This consistency is particularly evident when compared to Quantile Regression's variability of 0.147, indicating that DeepSupp provides more reliable performance across varying market conditions and time periods.

\begin{figure*}[h!]
  \centering
  \includegraphics[width=0.8\textwidth]{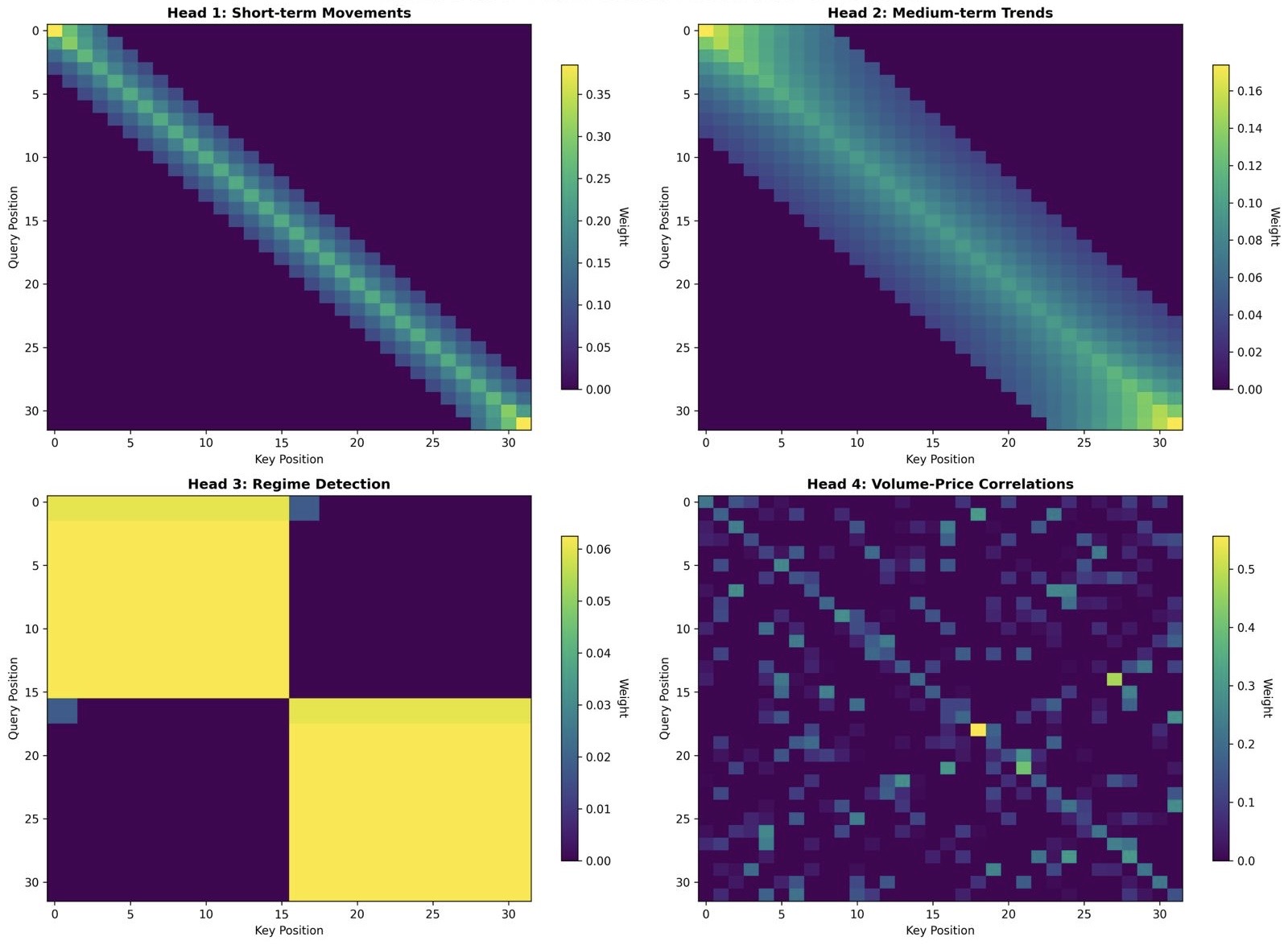}
  \caption{DeepSupp Mult-Head Attention Overview.}
  \label{fig:attention_overview}
\end{figure*}

The progression from heads one to four in Figure 3 shows an evolution from highly structured linear patterns to increasingly complex patterns. Head one follows a simple mathematical function, head two follows the same function with different parameters, head three introduces discrete state changes, and head four introduces apparent randomness. Such behavior suggests the model has learned a hierarchy of temporal patterns starting with simple local dependencies, progressing to longer range dependencies, then to regime based dependencies, and finally to exception based dependencies.

The attention weights themselves show different statistical distributions across heads. Heads one and two show smooth continuous distributions following exponential or Gaussian curves. Head three shows a bimodal distribution with values clustered near zero and one. Head four shows a power law distribution with many small values and few large spikes.

The spatial patterns reveal different types of market memory. The diagonal patterns indicate short term memory where recent events influence current decisions. The block patterns indicate regime memory where events within the same market phase influence each other regardless of temporal distance. The sparse patterns indicate crisis memory where rare but extreme events create long lasting influence patterns.

Such patterns suggest that DeepSupp has automatically discovered the fundamental structure of market behavior without being explicitly programmed with financial theory. The linear patterns correspond to momentum and mean reversion effects, while the block patterns correspond to market regime changes like bull and bear markets. The complex patterns demonstrate tail risk events and black swan phenomena that traditional models struggle to capture.

In Figure 4, we show performance comparison of DeepSupp and a popular Moving Average-based methods. The Moving Average Method demonstrates a clear limitation in its approach to support level identification through the tendency to generate closely spaced support levels that cluster around similar price ranges. As observed in the chart, seven support levels (ranging from approximately \$135 to \$143) are tightly packed with minimal separation between consecutive levels. This clustering effect results in redundant signals that provide little differentiation between support zones, causing the miss of critical support breaks or fail to identify the most significant support levels.

In contrast, DeepSupp exhibits a more strategic approach to support level placement by creating varied gaps between identified support levels, effectively highlighting distinct support zones with different significance levels. The method shows the ability to differentiate between major and minor support areas. Such varied placement suggests that DeepSupp employs sophisticated ideas to assess the relative importance of different price levels, creating a hierarchical support structure that better reflects actual market dynamics. The larger gaps between levels indicate that the method filters out noise and focuses on identifying support zones that represent genuine areas of buying interest or technical significance.

\begin{figure*}[h!]
  \centering
  \includegraphics[width=0.8\textwidth]{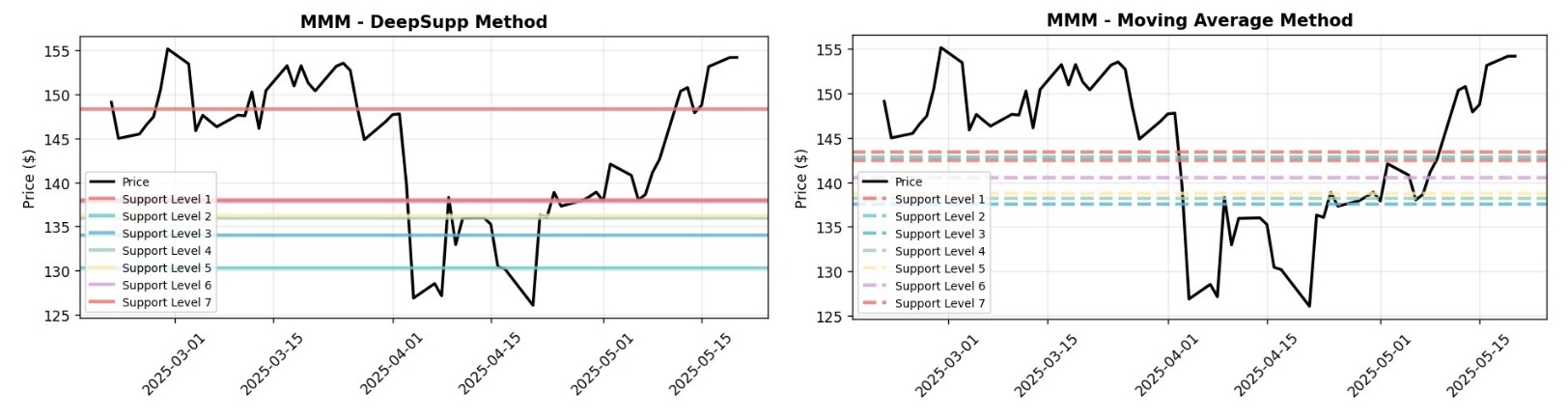}
  \caption{Performance Comparison: DeepSupp \& Moving Average Method.}
  \label{fig:performance_comparison}
\end{figure*}

The comparative empirical analysis demonstrates DeepSupp's superior performance relative to traditional moving average approaches, establishing a foundation for broader methodological comparisons in support level identification research. While the Moving Average Method represents one baseline approach among several conventional techniques, DeepSupp's varied level placement and dynamic adaptation capabilities highlight the advantages of our proposed approach.

\section{Conclusion}

In this study, we introduced DeepSupp, an attention-driven correlation–pattern framework for dynamic support-level identification in equity markets. DeepSupp combines rolling Spearman correlation matrices that capture evolving price–volume relations, a permutation-invariant multi-head attention auto-encoder for compact representation learning, and a density-based clustering stage that converts latent embeddings into discrete support zones.

The experiments across S\&P 500 tickers confirm our approach’s effectiveness, achieving the highest overall score among all tested methods. Although traditional methods like Local Minima excelled in isolated metrics, the balanced performance of DeepSupp in all evaluation dimensions establishes it as the most reliable solution for practical applications.

DeepSupp's core strength lies in its dynamic correlation analysis using sliding windows and multi-head attention, which captures evolving market relationships that static methods miss. The integration of DBSCAN clustering automatically extracts support levels from learned embeddings without predefined thresholds, identifying dense regions that correspond to different market regimes while maintaining robustness to noise.

Future research directions include experiment basis enlargement, evaluation of less liquid equities, fixed-income products, and cryptocurrencies to confirm portability, as well as inference time speedup for easier practical deployment scenarios.

Overall, DeepSupp demonstrates that attention-based correlation analysis offers a reliable and scalable approach to support level identification, effectively bridging traditional technical analysis and modern machine-learning methods.

\bibliographystyle{ieee_fullname}
\bibliography{egbib}
\end{document}